\documentclass[preprint]{revtex4-1} 

\usepackage{amsfonts} 
\usepackage{graphicx}
\usepackage{amsmath, amssymb}
\usepackage{slashed}
\usepackage{braket}
\usepackage{physics}
\usepackage[bottom]{footmisc}
\graphicspath{  {images/} }
\begin{document}


\title{Algebra for Fractional Statistics - interpolating from fermions to bosons}

\author{Satish Ramakrishna}
\email{ramakrishna@physics.rutgers.edu}
\affiliation{Department of Physics \& Astronomy, Rutgers, The State University of New Jersey, 136 Frelinghuysen Road
Piscataway, NJ 08854-8019}


\date{\today}

\begin{abstract}

This article constructs the Hilbert space for the algebra  $\alpha \beta - e^{i \theta} \beta \alpha = 1 $  that provides a continuous interpolation between the Clifford and Heisenberg algebras. This particular form is inspired by the properties of anyons. We study the eigenvalues of a generalized number operator (${\cal N} = \beta \alpha$) and construct the Hilbert space, classified by values of a complex coordinate ($\lambda_0$): the eigenvalues lie on a circle.  For $\theta$ being an irrational multiple of $2 \pi$, we get an infinite-dimensional representation, however for a rational multiple ($\frac{M}{N}$) of $2 \pi$, it is finite-dimensional, parametrized by the complex coordinate $\lambda_0$.  The case for $N=2 \: ; \: \theta=\pi$ is the usual Clifford algebra for fermions, while the case for $N=\infty \: ; \: \theta=0$ is the Heisenberg algebra of bosons, albeit with two copies for positive and negative eigenvalues. We find a smooth transition from the fermion to the boson situation as $N \rightarrow \infty$ from $N=2$.  After constructing the Hilbert space from the algebra, the cases for $N=2,3$ can be mapped to $SU(2)$. Then, we motivate the study of coherent states, rather generally. The coherent states are eigenstates of $\alpha$, the annihilation operator and are labeled by complex numbers for non-zero $\lambda_0$.

\end{abstract}

\maketitle 

The notion of fractional or generalized statistics has been studied intensively from many different directions. There are three main approaches. The oldest is the study of q-commutators (or q-statistics), which studied deviations from the Pauli principle by studying small deviations from the basic anti-commutator for fermions. The second is the study of q-deformed oscillators, which studies the  deformation of the harmonic oscillator algebra under a rather general scheme. This second approach is further differentiated into whether we consider the non-bosonic states to be only singly-occupied or not. We approach this topic from the point of view of interpolating between fermions and bosons, with an aim to encompass fractional statistics, with variable occupancy.

We start with the following bracket relation, which we refer to as a $\theta$-commutator
\begin{eqnarray}
\alpha \beta - e^{i \theta} \beta \alpha \equiv \bigg( \alpha, \beta \bigg)_{\theta} = 1 
\end{eqnarray}
Formally, this is a special case of q-on (or q-deformed) algebra, for $q = e^{i \theta}$, while the right-hand-side is 1 instead of the usual $q^N$ for a q-deformed algebra.  This may be regarded as an interpolation between the commutator for bosons and the anti-commutator for fermions. By analogy with the algebra for bosons and fermions, the operator $\alpha$ may be taken as the annihilation operator for the algebra, while $\beta$ is the creation operator, though we could switch the identifications and derive an entirely similar algebra (see Appendix 1). We do not assume that $\alpha$ and $\beta$ are hermitian conjugates of each other.

This algebra is also inspired by the braiding requirements for anyonic variables, where we can braid variables ``over'' and ``under'' each other, with factors $e^{\pm i \theta}$. We have chosen to work with the plus sign in the above; accordingly, we will use the notation $z = e^{i \theta}$.
We have two classes of possibilities for $\theta$. $\theta$ could be a rational multiple of $2 \pi$, i.e., of the form $\theta = \frac{2 \pi M}{N}$ where $M, N$ are co-prime non-zero natural numbers: in full generality, we could assume that $M<N$. Alternatively, $\theta$ could be an irrational multiple of $2 \pi$. We will primarily study the rational multiple case, but make comments where necessary about the other (irrational multiple) case. 
Note that with $M=1$ and $N=2$, the expression above reduces to a fermionic anti-commutator while for $N=\infty$, to a bosonic commutator.

For general integral $N$, we begin by thinking of $\alpha, \beta$ as ordinary matrices and construct the vector space they operate on. As we will discover, the results will indeed satisfy the axioms for a Hilbert space - the defining eigenvectors will constitute a complete vector space and we can define the usual inner product with positive norm.

We start by considering the operator ${\cal N} = \beta \alpha$. The following bracket relations can be quickly verified
\begin{eqnarray}
\bigg( \alpha, {\cal N} \bigg)_{\theta} = \alpha  \\
\bigg( {\cal N}, \beta \bigg)_{\theta} = \beta
\end{eqnarray}

These relations imply
\begin{eqnarray}
{\cal N} \alpha = \alpha \frac{{\cal N} -1}{z} \nonumber \\
{\cal N} \beta = \beta (1+z {\cal N})
\end{eqnarray}
Suppose $\vec \phi_a$ were an eigenvector of the operator $\cal N$, with eigenvalue $s_a$. If we were to operate on $\vec \phi_a$ with $\alpha$, using the above relations, we would get an eigenvector of $\cal N$ with eigenvalue $\frac{s_a -1}{z}$. We could label this state $\vec \phi_{a-1}$. Conversely, $\beta$ takes the eigenvector $\vec \phi_{a-1}$ and produces an eigenvector of $\cal N$,  presumably $\vec \phi_a$, with eigenvalue $s_a$. In this sense, we can think of $\alpha$ and $\beta$ as decrementing and incrementing operators. 

In fact, the recursion relation for the eigenvalues of the operator $\cal N$ is 
\begin{eqnarray}
s_a = 1 + z s_{a-1}
\end{eqnarray}

Next, we reason in the following manner. We define the real and imaginary parts of ${\cal N}$, i.e., ${\cal N}_R, {\cal N}_I$ and require that $[{\cal N}_R, {\cal N}_I]=0$. This immediately implies that $[{\cal N},{\cal N}^{\dagger}]=0$, which condition is the definition of a "normal" matrix, {\it viz.} ${\cal N}$ is a normal matrix. If ${\cal N}$ is normal, then ${\cal N}, {\cal N}^{\dagger}$ and the combinations ${\cal N}{\cal N}^{\dagger}$ and ${\cal N}^{\dagger}{\cal N}$ all commute with each other, i.e., they share the same eigenvectors.

Since ${\cal N}^{\dagger} {\cal N}$ and ${\cal N}{\cal N}^{\dagger}$ are both hermitian, their eigenvectors form an orthonormal, complete set. In this orthonormal basis, since $\alpha$ is a decrementing operator, while $\beta$ is an incrementing operator,  we immediately deduce that $\alpha$ is an upper-diagonal matrix, while $\beta$ is a lower-diagonal matrix. Additionally, it is possible to arrange things so that $\alpha$ and $\beta$ are transposes of each other, though this is not the only possibility (see the text after Equation (9)).

One consequence of the above is that any operator in this space can be written as a polynomial with powers of $\alpha$ and $\beta$. This is because any operator will take states in this complete, orthonormal basis to linear combinations of the basis vectors. Such a linear combination of states can be formed by repeatedly applying powers of $\alpha$ and $\beta$ to a starting state such as $\vec \phi_a$. In addition, only specific types of combinations of $\alpha$ and $\beta$ commute with $\cal N$, as discussed in Appendix 4.

Suppose we start with an eigenvector with eigenvalue (for the operator $\cal N$) of $\lambda_0$, with $\theta= 2 \pi \frac{M}{N}$. Applying $\beta$ successively to this eigenvector produces new eigenvectors with eigenvalues: $1+z \lambda_0, \: 1+z+z^2 \lambda_0, \: 1+z+z^2+z^3 \lambda_0, ..., ,\: 1+z+z^2+...+z^{N-2}+z^{N-1} \lambda_0, \: 1+z+z^2+...+z^{N-1}+z^N \lambda_0, ...$. With $\theta$ of the form $2 \pi \frac{M}{N}$, then $1+z+z^2+...+z^{N-1}=0$ and $z^N=1$, so that the sequence of eigenvalues repeats after $N-1$ applications of $\alpha$ to the starting eigenvector with eigenvalue $\lambda_0$. There are thus $N$ independent eigenvalues. Also, we note that $1+z \lambda_0 = \frac{\lambda_0 - (1+z+...+z^{N-2})}{z^{N-1}}$ etc., so that the sequence of eigenvalues of $\beta$ are the same as the eigenvalues of $\alpha$, except they are traversed in the opposite direction upon repeated application of $\beta$. This is how we would expect decrementing and incrementing operators to work on the eigenvectors.

In addition, the actual eigenvalues for the case $M=1$ as well as the case $M>1$, with $M, N$ co-prime are the same, just arrived at in different order. Hence, we could, in all generality, assume $M=1$ (the $\alpha$ and $\beta$ are the same matrices if $M>1$, just with permuted rows).

If $\theta$ were an irrational multiple of $2 \pi$, the eigenvalues would never repeat upon repeated applications of $\alpha$ or $\beta$, yet, since the relations $\lambda_0 \rightarrow \lambda_1 = (1+z \lambda_0)$ and $\lambda_1 \rightarrow \lambda_0 = \frac{\lambda_1-1}{z}$, the set of eigenvalues of $\alpha$ and $\beta$ are the same, though infinite in number. Summarizing, the matrix representation of $\alpha$ and $\beta$ are $N$-dimensional for $\theta = 2 \pi \frac{M}{N}$ and infinite-dimensional for $\theta$ an irrational multiple of $2 \pi$.

These eigenvalues sit on the same  circle in the complex plane, for all $\theta$. For what follows, however, we specialize to the case where $\theta$ is a rational multiple of $2 \pi$ and also set $M=1$. Then we construct the matrix representations of $\alpha$ and $\beta$ in the eigen basis of $\cal N$.

The sequence of eigenvalue points on the complex plane is $\lambda_0$, followed by the successive eigenvalues $1+\lambda_0 z, q+z+\lambda_0 z^2, q+z+z^2+\lambda_0 z^3, ..., 1+z+z^2+...+z^{N-2}+\lambda_0 z^{N-1}$ (where, since $M=1$, $z=e^{\frac{2 \pi i}{N}}$).

There are a few simple ways to show that these points lie on a circle. One method is as follows, a second is described in Appendix 2. The average of all the complex eigenvalues is ${\cal Z} = \frac{1}{1-z}$ (independent of $\lambda_0$), which is the complex coordinate for the center. Then, the magnitude of the difference between each of the points in the above sequence and this central point is ${\cal R} = |\frac{1+(z-1) \times \lambda_0}{1-z}|$, i.e., each eigenvalue point is the same distance $|\frac{1+(z-1) \times \lambda_0}{1-z}|$ from the center (see Appendix 2). This can then be identified as the radius of the circle.

Explicitly, the average of the points is ${\cal O} $
\begin{eqnarray}
=\frac{(\lambda_0) + (1+z \lambda_0) + (1+z+z^2 \lambda_0) + (1+z+z^2+z^3 \lambda_0) + ... + (1+z+z^2+...+z^{N-2}+z^{N-1} \lambda_0)}{N} \: \: \: \: \: \: \: \: \: \:  \nonumber \\
= \frac{\lambda_0 (1+z+z^2+...+z^{N-1}) + (N-1) z^0 + (N-2) z^1 + (N-3) z^2 + ... + (N-[N-1]) z^{N-2} }{N}  \: \: \: \: \: \: \: \: \: \: \: \: \: \: \: \: \: \: \:  \nonumber \\
=\frac{0 + (N-1) z^0 + (N-2) z^1 + (N-3) z^2 + ... + (N-[N-1]) z^{N-2} }{N}  \: \: \: \: \: \: \: \: \: \: \: \: \: \: \: \: \: \: \:  \: \: \: \: \: \: \: \: \: \: \: \: \: \: \: \: \: \: \:  \nonumber \\
= \frac{1}{1-z} = \frac{i e^{- i \frac{\theta}{2}}}{2 \sin \frac{\theta}{2}} = \frac{1}{2} + \frac{i}{2 \tan \frac{\theta}{2}} \: \: \: \: \: \: \: \: \: \: \: \: \: \: \: \: \: \: \: \: \: \: \: \: \: \: \: \: \: \: \: \: \: \: \: \: \: \: \: \: \: \: \: \: \: \: \: \: \: \: \: \: \nonumber
\end{eqnarray}
i.e., the center of the eigenvalue circle is {\bf {exactly}} at the same spot for all values of the starting eigenvalue $\lambda_0$. The radius does depend on $\lambda_0$, it is the magnitude of the difference between any eigenvalue and the complex coordinate of the center.

The angle between the lines connecting the center $\frac{1}{1-z}$ and successive points $1+z+...+z^{p-2}+\lambda_0 z^{p-1}$ and $1+z+z^2+...+z^{p-1}+\lambda_0 z^p$ is easily computed by the ratio of the complex differences to the center - it is $z$, i.e., the angle between the lines is $\theta$. 

From the above, it is clear that we can label the representations (for a given $N$) by either the complex number $\lambda_0$ (two real numbers) or the two real numbers representing the radius of the eigenvalue circle ${\cal R}$ and the polar angle $\chi \in (0,\frac{2 \pi}{N})$.

If $\lambda_0=0$, then the first point in the above sequence is $0$ and it is (one of the) lowest point(s) on the circle. When we consider non-zero $\lambda_0$, that lowest point is mapped to a  rotated point on a bigger circle (radius $|1+(z-1)\lambda_0|$ with the same center. It is easily checked that the polar angle of the rotated point is bigger than that of the initial point by  $\alpha = \frac{1}{i} \log \frac{1+(z-1)\lambda_0}{|1+(z-1)\lambda_0|}$. This is easily proved to be true for every eigenvalue in the $\lambda_0=0$ circle - it is rotated by the above angle when moved to the larger circle for $\lambda_0 \ne 0$.

In addition, the magnitude of the separation distance between successive eigenvalues on the circle is $|1+(z-1)\lambda_0|$.

\begin{figure}[h!]
\caption{Eigenvalue Circle, M=1, N=100, $\lambda_0$=0, as well as $\lambda_0=15+3i$}
\centering
\includegraphics[scale=.36]{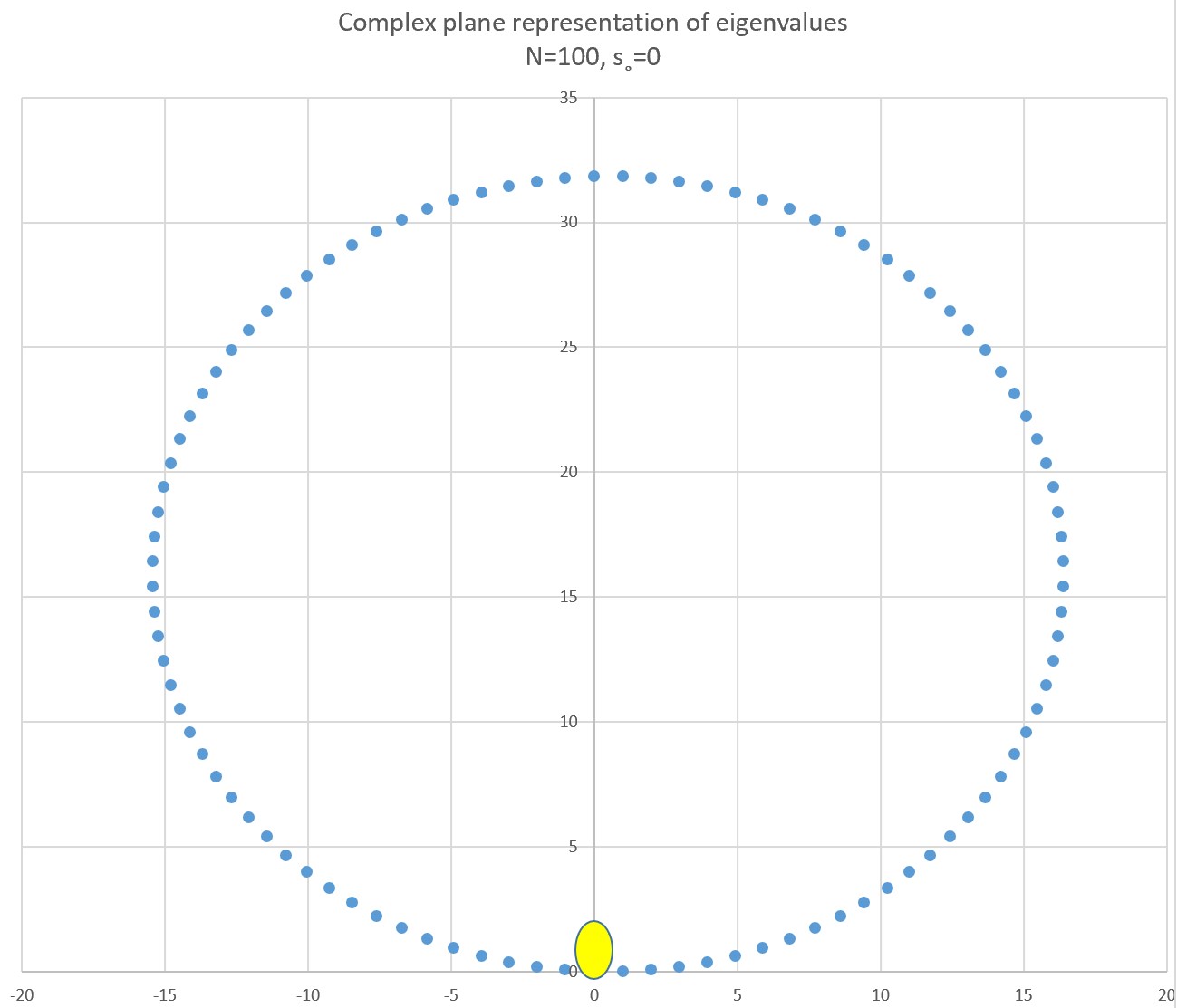}
\includegraphics[scale=0.45]{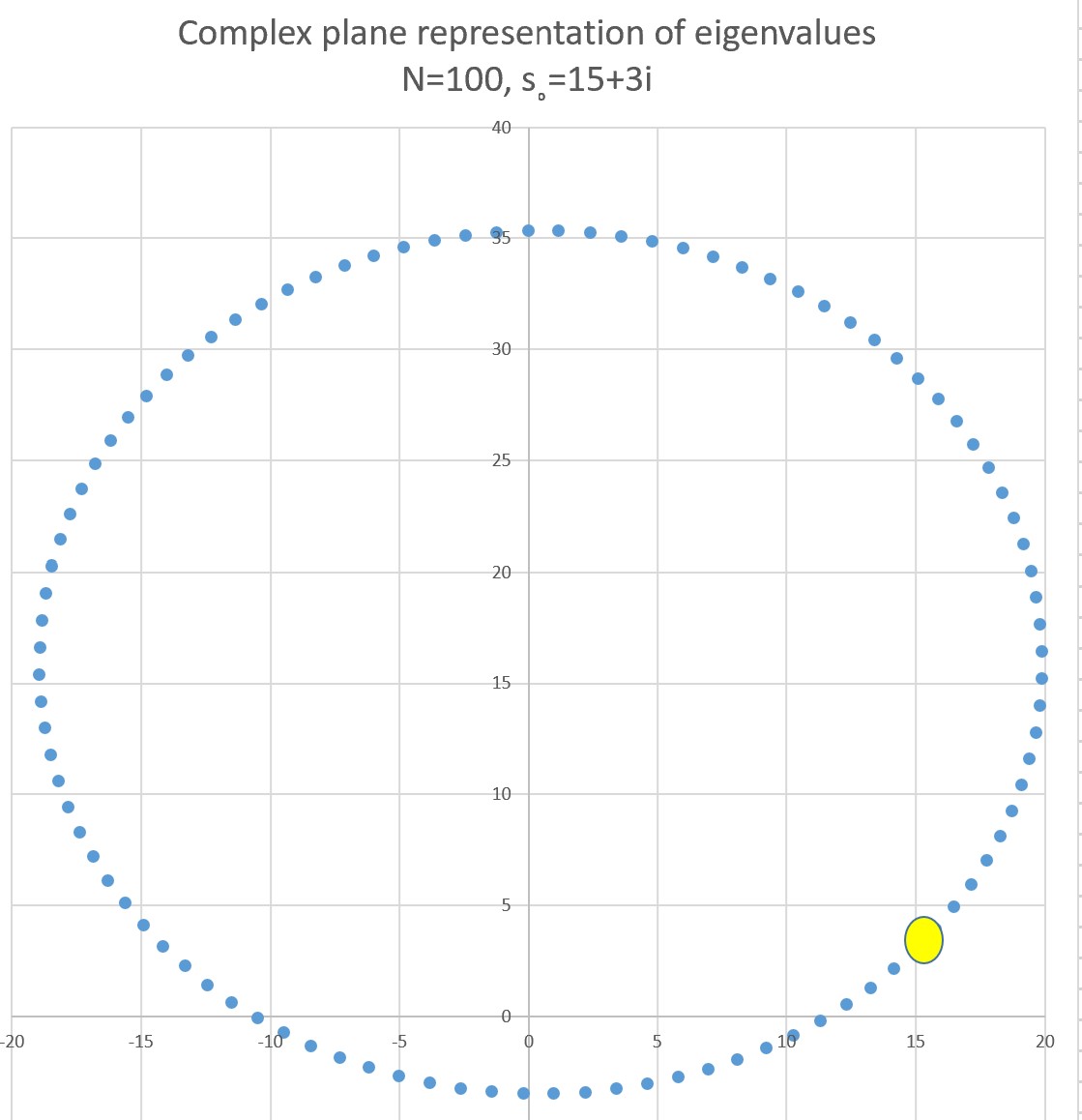}
\end{figure}

Since $\alpha$ ``decrements'' states, it is an $N \times N$ matrix with upper diagonal non-zero terms only (as well as one in the bottom left corner to cycle through states again), in the basis where ${\cal N} = \beta \alpha$ is diagonal. Let's state that $\beta \alpha$ is diagonal, with eigenvalues denoted by, say $\lambda_i$, $i=0...(N-1)$. Hence, purely by inspection and the expected upper-diagonal structure of $\alpha$, as well as requiring that we ``split'' the eigenvalues to make $\beta$ and $\alpha$ transposes of each other,  it must have $\sqrt{\lambda_{i+1}}$ as elements. Hence the matrix $\alpha^{\dagger} \alpha$, by inspection, must have diagonal elements $|\lambda_i|$, $0=1...(N-1)$. Also by inspection, the matrices $\alpha \beta$ and $\alpha \alpha^{\dagger}$ are diagonal; in fact, the elements of $\alpha \beta$ must be $\lambda_{i+1}$, $i=0...(N-1)$, where $\lambda_N=\lambda_0$. This additionally implies that the matrix $\alpha \alpha^{\dagger}$ must have diagonal elements $|\lambda_{i+1}|$, $i=0...(N-1)$.

Visually,
\begin{eqnarray}
\beta \alpha=
\left(\begin{array}{cccccccc} {\bf \lambda_0} & \: \: \: 0 &\: \: \:  0 &\: \: \:  0 &\: \: \:  0 & ... & \: \: \:  0 &0\\
					     0 & \: \: \: {\bf \lambda_1} & \: \: \: 0 & \: \: \:  0 & \: \: \:  0 & ... & \: \: \:  0 &0 \\
					     0 & \: \: \: 0 & \: \: \: {\bf \lambda_2} &\: \: \: 0 & \: \: \:  0 &... & \: \: \:  0 &0 \\
					     0 & \: \: \: 0 & \: \: \: 0 &\: \: \: {\bf \lambda_3} & \: \: \: 0 &... & \: \: \:  0 &0 \\
					     . \\
					     . \\
					     . \\
					     0 & \: \: \: 0 & \: \: \: 0 &\: \: \:  0 & \: \: \:  0 & ... &\: \: \:  {\bf \lambda_{N-2}}  &0  \\
					    0 & \: \: \: 0 & \: \: \: 0 &\: \: \:  0 & \: \: \:  0 & ... &\: \: \:  0 & {\bf \lambda_{N-1}} \end{array}\right) \nonumber
\end{eqnarray}
\begin{eqnarray}
\alpha \beta=
\left(\begin{array}{cccccccc} {\bf \lambda_1} & \: \: \: 0 &\: \: \:  0 &\: \: \:  0 &\: \: \:  0 & ... & \: \: \:  0 &0\\
					     0 & \: \: \: {\bf \lambda_2} & \: \: \: 0 & \: \: \:  0 & \: \: \:  0 & ... & \: \: \:  0 &0 \\
					     0 & \: \: \: 0 & \: \: \: {\bf \lambda_3} &\: \: \: 0 & \: \: \:  0 &... & \: \: \:  0 &0 \\
					     0 & \: \: \: 0 & \: \: \: 0 &\: \: \: {\bf \lambda_4} & \: \: \: 0 &... & \: \: \:  0 &0 \\
					     . \\
					     . \\
					     . \\
					     0 & \: \: \: 0 & \: \: \: 0 &\: \: \:  0 & \: \: \:  0 & ... &\: \: \:  {\bf \lambda_{N-1}}  &0  \\
					    0 & \: \: \: 0 & \: \: \: 0 &\: \: \:  0 & \: \: \:  0 & ... &\: \: \:  0 & {\bf \lambda_{0}} \end{array}\right) \nonumber
\end{eqnarray}
\begin{eqnarray}
\alpha=
\left(\begin{array}{cccccccc} {\bf 0} & \: \: \: \sqrt{\lambda_1} &\: \: \:  0 &\: \: \:  0 &\: \: \:  0 & ... & \: \: \:  0 &0\\
					     0 & \: \: \: {\bf 0} & \: \: \:\sqrt{\lambda_2} & \: \: \:  0 & \: \: \:  0 & ... & \: \: \:  0 &0 \\
					     0 & \: \: \: 0 & \: \: \: {\bf 0} &\: \: \: \sqrt{\lambda_3} & \: \: \:  0 &... & \: \: \:  0 &0 \\
					     0 & \: \: \: 0 & \: \: \: 0 &\: \: \: {\bf 0} & \: \: \:  \sqrt{\lambda_4} &... & \: \: \:  0 &0 \\
					     . \\
					     . \\
					     . \\
					     0 & \: \: \: 0 & \: \: \: 0 &\: \: \:  0 & \: \: \:  0 & ... &\: \: \:  {\bf 0}  & \sqrt{\lambda_{N-1}}  \\
					    \sqrt{\lambda_0} & \: \: \: 0 & \: \: \: 0 &\: \: \:  0 & \: \: \:  0 & ... &\: \: \:  0 & {\bf 0} \end{array}\right) \nonumber
\end{eqnarray}
\begin{eqnarray}
					     \beta = \alpha^T =
\left(\begin{array}{cccccccc} {\bf 0} & \: \: \: 0 &\: \: \:  0 &\: \: \:  0 & \: \: \:  0 & ... & \: \: \:  0 &\sqrt{\lambda_0}\\
					    \sqrt{\lambda_1}& \: \: \: {\bf 0} & \: \: \: 0 & \: \: \:  0 & \: \: \:  0 & ... &\: \: \:  0 & 0 \\
					     0 & \: \: \:\sqrt{\lambda_2} & \: \: \: {\bf 0} &\: \: \: 0 & \: \: \:  0 & ... &\: \: \:  0 & 0 \\
					     0 & \: \: \: 0 & \: \: \:\sqrt{\lambda_3}& \: \: \: {\bf 0}& \: \: \:  0 & ... &\: \: \:  0 & 0\\
					      0 & \: \: \: 0 & \: \: \: 0 &\: \: \: \sqrt{\lambda_4} & \: \: \: {\bf 0}&  ... &\: \: \:  0 & 0\\
					     . \\
					     . \\
					     . \\
					     0& \: \: \: 0 & \: \: \: 0 &\: \: \:  0 & \: \: \:  0 & ... & \sqrt{\lambda_{N-1}} &  {\bf 0} \end{array}\right) \nonumber
\end{eqnarray}
\begin{eqnarray}
\alpha^*=
\left(\begin{array}{cccccccc} {\bf 0} & \: \: \: \sqrt{\lambda_1^*} &\: \: \:  0 &\: \: \:  0 &\: \: \:  0 & ... & \: \: \:  0 &0\\
					     0 & \: \: \: {\bf 0} & \: \: \:\sqrt{\lambda_2^*} & \: \: \:  0 & \: \: \:  0 & ... & \: \: \:  0 &0 \\
					     0 & \: \: \: 0 & \: \: \: {\bf 0} &\: \: \: \sqrt{\lambda_3^*} & \: \: \:  0 &... & \: \: \:  0 &0 \\
					     0 & \: \: \: 0 & \: \: \: 0 &\: \: \: {\bf 0} & \: \: \:  \sqrt{\lambda_4^*} &... & \: \: \:  0 &0 \\
					     . \\
					     . \\
					     . \\
					     0 & \: \: \: 0 & \: \: \: 0 &\: \: \:  0 & \: \: \:  0 & ... &\: \: \:  {\bf 0}  & \sqrt{\lambda_{N-1}^*}  \\
					    \sqrt{\lambda_0^*} & \: \: \: 0 & \: \: \: 0 &\: \: \:  0 & \: \: \:  0 & ... &\: \: \:  0 & {\bf 0} \end{array}\right) \nonumber
\end{eqnarray}
\begin{eqnarray}
					     \beta^* = \alpha^{\dagger} =
\left(\begin{array}{cccccccc} {\bf 0} & \: \: \: 0 &\: \: \:  0 &\: \: \:  0 & \: \: \:  0 & ... & \: \: \:  0 &\sqrt{\lambda_0^*}\\
					    \sqrt{\lambda_1^*}& \: \: \: {\bf 0} & \: \: \: 0 & \: \: \:  0 & \: \: \:  0 & ... &\: \: \:  0 & 0 \\
					     0 & \: \: \:\sqrt{\lambda_2^*} & \: \: \: {\bf 0} &\: \: \: 0 & \: \: \:  0 & ... &\: \: \:  0 & 0 \\
					     0 & \: \: \: 0 & \: \: \:\sqrt{\lambda_3^*}& \: \: \: {\bf 0}& \: \: \:  0 & ... &\: \: \:  0 & 0\\
					      0 & \: \: \: 0 & \: \: \: 0 &\: \: \: \sqrt{\lambda_4^*} & \: \: \: {\bf 0}&  ... &\: \: \:  0 & 0\\
					     . \\
					     . \\
					     . \\
					     0& \: \: \: 0 & \: \: \: 0 &\: \: \:  0 & \: \: \:  0 & ... & \sqrt{\lambda_{N-1}^*} &  {\bf 0} \end{array}\right) \nonumber
\end{eqnarray}
\begin{eqnarray}
\alpha^{\dagger} \alpha=
\left(\begin{array}{cccccccc} {\bf |\lambda_0|} & \: \: \: 0 &\: \: \:  0 &\: \: \:  0 &\: \: \:  0 & ... & \: \: \:  0 &0\\
					     0 & \: \: \: {\bf |\lambda_1|} & \: \: \: 0 & \: \: \:  0 & \: \: \:  0 & ... & \: \: \:  0 &0 \\
					     0 & \: \: \: 0 & \: \: \: {\bf |\lambda_2|} &\: \: \: 0 & \: \: \:  0 &... & \: \: \:  0 &0 \\
					     0 & \: \: \: 0 & \: \: \: 0 &\: \: \: {\bf |\lambda_3|} & \: \: \: 0 &... & \: \: \:  0 &0 \\
					     . \\
					     . \\
					     . \\
					     0 & \: \: \: 0 & \: \: \: 0 &\: \: \:  0 & \: \: \:  0 & ... &\: \: \:  {\bf |\lambda_{N-2}|}  &0  \\
					    0 & \: \: \: 0 & \: \: \: 0 &\: \: \:  0 & \: \: \:  0 & ... &\: \: \:  0 & {\bf |\lambda_{N-1}|} \end{array}\right) \nonumber
\end{eqnarray}
\begin{eqnarray}
\alpha \alpha^{\dagger}=
\left(\begin{array}{cccccccc} {\bf |\lambda_1|} & \: \: \: 0 &\: \: \:  0 &\: \: \:  0 &\: \: \:  0 & ... & \: \: \:  0 &0\\
					     0 & \: \: \: {\bf |\lambda_2|} & \: \: \: 0 & \: \: \:  0 & \: \: \:  0 & ... & \: \: \:  0 &0 \\
					     0 & \: \: \: 0 & \: \: \: {\bf |\lambda_3|} &\: \: \: 0 & \: \: \:  0 &... & \: \: \:  0 &0 \\
					     0 & \: \: \: 0 & \: \: \: 0 &\: \: \: {\bf |\lambda_4|} & \: \: \: 0 &... & \: \: \:  0 &0 \\
					     . \\
					     . \\
					     . \\
					     0 & \: \: \: 0 & \: \: \: 0 &\: \: \:  0 & \: \: \:  0 & ... &\: \: \:  {\bf |\lambda_{N-1}|}  &0  \\
					    0 & \: \: \: 0 & \: \: \: 0 &\: \: \:  0 & \: \: \:  0 & ... &\: \: \:  0 & {\bf |\lambda_{0}|} \end{array}\right) \nonumber
\end{eqnarray}
We can now  write down regular and $\theta$ commutators for $\alpha$ and $\alpha^{\dagger}$ too, as
\begin{eqnarray}
\alpha \beta -\beta \alpha =
\left(\begin{array}{cccccccc} {\bf \lambda_1 - \lambda_0} & \: \: \: 0 &\: \: \:  0 &\: \: \:  0 &\: \: \:  0 & ... & \: \: \:  0 &0\\
					     0 & \: \: \: {\bf \lambda_2 - \lambda_1} & \: \: \: 0 & \: \: \:  0 & \: \: \:  0 & ... & \: \: \:  0 &0 \\
					     0 & \: \: \: 0 & \: \: \: {\bf \lambda_3 - \lambda_2} &\: \: \: 0 & \: \: \:  0 &... & \: \: \:  0 &0 \\
					     0 & \: \: \: 0 & \: \: \: 0 &\: \: \: {\bf \lambda_4 - \lambda_3} & \: \: \: 0 &... & \: \: \:  0 &0 \\
					     . \\
					     . \\
					     . \\
					     0 & \: \: \: 0 & \: \: \: 0 &\: \: \:  0 & \: \: \:  0 & ... &\: \: \:  {\bf \lambda_{N-1}-\lambda_{N-2}}  &0  \\
					    0 & \: \: \: 0 & \: \: \: 0 &\: \: \:  0 & \: \: \:  0 & ... &\: \: \:  0 & {\bf \lambda_0- \lambda_{N-1}} \end{array}\right) \nonumber
\end{eqnarray}
and
\begin{eqnarray}
\alpha \alpha^{\dagger} - \alpha^{\dagger} \alpha  \: \: \: \: \: \: \: \: \: \:  \: \: \: \: \: \: \: \: \: \:  \: \: \: \: \: \: \: \: \: \:  \: \: \: \: \: \: \: \: \: \:  \: \: \: \: \: \: \: \: \: \:  \: \: \: \: \: \: \: \: \: \:    \: \: \: \: \: \: \: \: \: \:  \: \: \: \: \: \: \: \: \: \: \nonumber \\
= \left(\begin{array}{cccccccc} {\bf |\lambda_1| - |\lambda_0|} & \: \: \: 0 &\: \: \:  0 &\: \: \:  0 &\: \: \:  0 & ... & \: \: \:  0 &0\\
					     0 & \: \: \: {\bf |\lambda_2| - |\lambda_1|} & \: \: \: 0 & \: \: \:  0 & \: \: \:  0 & ... & \: \: \:  0 &0 \\
					     0 & \: \: \: 0 & \: \: \: {\bf |\lambda_3| - |\lambda_2|} &\: \: \: 0 & \: \: \:  0 &... & \: \: \:  0 &0 \\
					     0 & \: \: \: 0 & \: \: \: 0 &\: \: \: {\bf |\lambda_4| - |\lambda_3|} & \: \: \: 0 &... & \: \: \:  0 &0 \\
					     . \\
					     . \\
					     . \\
					     0 & \: \: \: 0 & \: \: \: 0 &\: \: \:  0 & \: \: \:  0 & ... &\: \: \:  {\bf |\lambda_{N-1}|-|\lambda_{N-2}|}  &0  \\
					    0 & \: \: \: 0 & \: \: \: 0 &\: \: \:  0 & \: \: \:  0 & ... &\: \: \:  0 & {\bf |\lambda_0|- |\lambda_{N-1}|} \end{array}\right) \nonumber
\end{eqnarray}
as well as
\begin{eqnarray}
\alpha \alpha^{\dagger} - {\cal C} \alpha^{\dagger} \alpha = 1 \rightarrow\: \: \: \: \: \: \: \: \: \:  \: \: \: \: \: \: \: \: \: \:  \: \: \: \: \: \: \: \: \: \:  \: \: \: \: \: \: \: \: \: \:  \: \: \: \: \: \: \: \: \: \:  \: \: \: \: \: \: \: \: \: \:    \: \: \: \: \: \: \: \: \: \:  \: \: \: \: \: \: \: \: \: \: \nonumber \\
{\cal C} = \left(\begin{array}{cccccccc} {\bf \frac{ |\lambda_1| - 1}{|\lambda_0|}} & \: \: \: 0 &\: \: \:  0 &\: \: \:  0 &\: \: \:  0 & ... & \: \: \:  0 &0\\
					     0 & \: \: \: {\bf \frac{ |\lambda_2| -1 }{|\lambda1|}} & \: \: \: 0 & \: \: \:  0 & \: \: \:  0 & ... & \: \: \:  0 &0 \\
					     0 & \: \: \: 0 & \: \: \: {\bf \frac{ |\lambda_3| - 1}{|\lambda_2|}} &\: \: \: 0 & \: \: \:  0 &... & \: \: \:  0 &0 \\
					     0 & \: \: \: 0 & \: \: \: 0 &\: \: \: {\bf \frac{ |\lambda_4| - 1}{|\lambda_3|}} & \: \: \: 0 &... & \: \: \:  0 &0 \\
					     . \\
					     . \\
					     . \\
					     0 & \: \: \: 0 & \: \: \: 0 &\: \: \:  0 & \: \: \:  0 & ... &\: \: \:  {\bf \frac{|\lambda_{N-1}|-1}{|\lambda_{N-2}|}}  &0  \\
					    0 & \: \: \: 0 & \: \: \: 0 &\: \: \:  0 & \: \: \:  0 & ... &\: \: \:  0 & {\bf \frac{ |\lambda_0|- 1}{|\lambda_{N-1}|}} \end{array}\right) \nonumber
\end{eqnarray}

Summarizing the results,
\begin{enumerate}
\item The eigenvalues of $\beta \alpha$ are $\lambda_i$
\item The eigenvalues of $\alpha^{\dagger} \alpha$ are $|\lambda_i|$.
\item ${\cal N}^{\dagger} = \alpha^{\dagger} \beta^{\dagger} = \alpha^{\dagger} \alpha^* = (\alpha_T \alpha)^*$, i.e., the operator $\alpha^{\dagger} \beta^{\dagger}$ is diagonal in the same basis and has eigenvalues $\lambda_i^*$.
\item The eigenbasis of $\beta \alpha$ is also the eigenbasis for $\alpha^{\dagger} \alpha$, hence the eigenvectors can be constructed to be orthonormal, as will be shown below. In addition, the eigenbasis forms a Hilbert space, since the norms of the eigenvectors are positive-definite.
\item The operators $\cal N = \beta \alpha$, ${\cal N}_R=\frac{{\cal N} + {\cal N}^{\dagger}}{2} = \frac{\beta \alpha +\alpha^{\dagger} \beta^{\dagger}}{2}$ and ${\cal N}_I=\frac{{\cal N} - {\cal N}^{\dagger}}{2i} = \frac{\beta \alpha-\alpha^{\dagger} \beta^{\dagger}}{2i}$ commute with each other and the eigenvalues of each are distinct. Hence, we can label the eigenvectors interchangeably by the eigenvalues of ${\cal N}$, ${\cal N}_R$ or ${\cal N}_I$.
\item In particular, since ${\cal N}_R$ and ${\cal N}_I$ are hermitian, the eigenvectors can be constructed to be orthonormal. In fact, noting the fact that the eigenvectors of a Hermitian matrix have positive norm, the space of eigenvectors satisfy all the axioms for a Hilbert space.
\end{enumerate}

\noindent The above visual argument can be made precise. We start with an eigenvector of $\cal N$, which is also an eigenstate of ${\cal N}_R$ and ${\cal N}_I$), eigenvalue $\lambda_0$ under $\cal N$ (eigenvalues $\frac{\lambda_0+\lambda_0^*}{2}$, $\frac{\lambda_0-\lambda_0^*}{2i}$ under ${\cal N}_R$, ${\cal N}_I$ respectively). Then, we derive all the successive eigenvectors and their eigenvalues. As proved in Appendix 3, the eigenvectors of ${\cal N}_R$, ${\cal N}_I$ and ${\cal N}$ are all the same and can be labeled simultaneously by the eigenvalues of the three operators, as $\frac{1}{2}(\lambda_i + \lambda^*_i)$, $\frac{1}{2i}(\lambda_i - \lambda^*_i)$ and $\lambda_i$. 
The eigenvalues we get by repeated applications of $\alpha$ are,
\begin{enumerate}
\item  for ${\cal N}$: $\lambda_0$, $\frac{\lambda_0-1}{z}$, $\frac{\lambda_0-(1+z)}{z^2}$, $\frac{\lambda_0-(1+z+z^2)}{z^3}$, ... $\frac{\lambda_0-(1+z+z^2+...+z^{N-2})}{z^{N-1}}$, $\frac{\lambda_0-(1+z+z^2+...+z^{N-1})}{z^{N}}, ... \: \: $.
\item  for ${\cal N}_R$:  Re $\lambda_0$, Re $\frac{\lambda_0-1}{z}$, Re $\frac{\lambda_0-(1+z)}{z^2}$, Re $\frac{\lambda_0-(1+z+z^2)}{z^3}$, ... Re $\frac{\lambda_0-(1+z+z^2+...+z^{N-2})}{z^{N-1}}$, Re $\frac{\lambda_0-(1+z+z^2+...+z^{N-1})}{z^{N}}, ... \: \: $.
\item  for ${\cal N}_I$:  Im $\lambda_0$,  Im $\frac{\lambda_0-1}{z}$,  Im $\frac{\lambda_0-(1+z)}{z^2}$, Im  $\frac{\lambda_0-(1+z+z^2)}{z^3}$, ...  Im $\frac{\lambda_0-(1+z+z^2+...+z^{N-2})}{z^{N-1}}$,  Im $\frac{\lambda_0-(1+z+z^2+...+z^{N-1})}{z^{N}}, ... \: \: $.
\end{enumerate}

Studying the above, since we already know that there are $N$ independent eigenvectors, the $N$ distinct eigenvalues of $\cal N$ are most appropriate to label the eigenvectors. The other commuting operators, {\it viz} ${\cal N}_R$ and ${\cal N}_I$ have degenerate eigenvalues, as is evident from an inspection of Fig. 2.

Hence, we define the normalized eigenstates of ${\cal N} = \beta \alpha$ as $\ket{\lambda_0}, \ket{\lambda_1}, \ket{\lambda_2},...,\ket{\lambda_{N-1}}$, where $\lambda_1=(1+z \lambda_0), \lambda_2=(1+z+z^2 \lambda_0), ... ,\lambda_{N-1}=1+z+z^2+...+z^{N-2}+z^{N-1} \lambda_0$ and equivalently $\lambda_0=\frac{\lambda_1-1}{z}, \lambda_1=\frac{\lambda_2-1}{z}, ..., \lambda_{N-1}=\frac{\lambda_0-1}{z}$.
Then, we construct the components of $\alpha$ and $\beta$.
We begin with the relationships (the $C$'s and $D$'s are normalization constants).

\begin{eqnarray}
\alpha \ket{\lambda_0} = C_0 \ket{\lambda_{N-1} \equiv \frac{\lambda_0-1}{z}} \nonumber \\
\alpha \ket{\lambda_{N-1}} = C_{N-1} \ket{s_{N-2}=\frac{\lambda_0 - (1+z)}{z^2} \equiv \frac{\lambda_0}{z^2}-\frac{1}{z}-\frac{1}{z^2}} \nonumber \\
. \nonumber \\
. \nonumber \\
. \nonumber \\
\alpha \ket{\lambda_0} = C_0 \ket{\lambda_{N-1}} \nonumber \\
\alpha \ket{\lambda_1} = C_1 \ket{\lambda_0} \nonumber \\
\alpha \ket{\lambda_2} = C_2 \ket{\lambda_1} \nonumber \\
. . . \nonumber \\
\beta \ket{\lambda_0} = D_0 \ket{\lambda_1 \equiv (1+z \lambda_0)} \nonumber
\end{eqnarray}
\begin{eqnarray}
\beta \ket{\lambda_1} = D_1 \ket{\lambda_2 \equiv (1+z+z^2 \lambda_0)} \nonumber \\
. \nonumber \\
. \nonumber \\
. \nonumber \\
\beta \ket{\lambda_{N-1}}= D_{N-1} \ket{\lambda_0}
\end{eqnarray}
For consistency with the results of applying the number operator ${\cal N}=\beta \alpha$ to these states, given that $\lambda_0, \lambda_1, ..., \lambda_{N-1}$ are the eigenvalues of the number operator, we get the following conditions,
\begin{eqnarray}
 C_0 \times D_{N-1} = \lambda_0 \nonumber \\
C_1 \times D_0 = \lambda_1 \nonumber \\
C_2 \times D_1 = \lambda_2 \nonumber \\
. \nonumber \\
. \nonumber \\
. \nonumber \\
C_{N-1} \times D_{N-2} = \lambda_{N-1} 
\end{eqnarray}
A natural choice (only one of several choices, though, as explained after Equation (9)) for the coefficients is
\begin{eqnarray}
C_0 = D_{N-1} = \sqrt{\lambda_0} \nonumber \\
C_1 = D_0 = \sqrt{\lambda_1} \nonumber \\
... \nonumber \\
C_{N-1} = D_{N-2} = \sqrt{\lambda_{N-1}}
\end{eqnarray}
This implies the matrix representations for the operators are

\begin{eqnarray}
\alpha=
\left(\begin{array}{cccccccc} {\bf 0} & \: \: \:  C_1 &\: \: \:  0 &\: \: \:  0 &\: \: \:  0 & ... & \: \: \:  0 &0\\
					     0 & \: \: \: {\bf 0} & \: \: \: C_2 & \: \: \:  0 & \: \: \:  0 & ... & \: \: \:  0 &0 \\
					     0 & \: \: \: 0 & \: \: \: {\bf 0} &\: \: \: C_3 & \: \: \:  0 &... & \: \: \:  0 &0 \\
					     0 & \: \: \: 0 & \: \: \: 0 &\: \: \: {\bf 0} & \: \: \:  C_4 &... & \: \: \:  0 &0 \\
					     . \\
					     . \\
					     . \\
					     0 & \: \: \: 0 & \: \: \: 0 &\: \: \:  0 & \: \: \:  0 & ... &\: \: \:  {\bf 0}  & C_{N-1}  \\
					     C_{0} & \: \: \: 0 & \: \: \: 0 &\: \: \:  0 & \: \: \:  0 & ... &\: \: \:  0 & {\bf 0} \end{array}\right) \: \: 
					     \beta = 
\left(\begin{array}{cccccccc} {\bf 0} & \: \: \: 0 &\: \: \:  0 &\: \: \:  0 & \: \: \:  0 & ... & \: \: \:  0 & D_{N-1}\\
					     D _0& \: \: \: {\bf 0} & \: \: \: 0 & \: \: \:  0 & \: \: \:  0 & ... &\: \: \:  0 & 0 \\
					     0 & \: \: \: D_1 & \: \: \: {\bf 0} &\: \: \: 0 & \: \: \:  0 & ... &\: \: \:  0 & 0 \\
					     0 & \: \: \: 0 & \: \: \: D_2& \: \: \: {\bf 0}& \: \: \:  0 & ... &\: \: \:  0 & 0\\
					      0 & \: \: \: 0 & \: \: \: 0 &\: \: \: D_3 & \: \: \: {\bf 0}&  ... &\: \: \:  0 & 0\\
					     . \\
					     . \\
					     . \\
					     0& \: \: \: 0 & \: \: \: 0 &\: \: \:  0 & \: \: \:  0 & ... & D_{N-2} &  {\bf 0} \end{array}\right) \nonumber
\end{eqnarray}
These matrices are transposes ($\beta =\alpha^T$) of each other, based on the choice made in Equation (8) and the discussions after Equation (9).

Additionally, we find 
\begin{eqnarray}
\beta^N=\alpha^N=C_0 \times C_1 \times ... \times C_{N-1} = D_0 \times D_1 \times ... \times D_{N-1} \nonumber \\
=\sqrt{\lambda_0 \times \lambda_1 \times \lambda_2 \times \lambda_3 ...\times \lambda_{N-1}}
\end{eqnarray}
For the special case $\lambda_0=0$, this is $0$.

It is not necessary for us to choose the conditions as in Equation (8), i.e., that $\beta = \alpha^T$. Suppose we chose to not pick this choice. In the basis where ${\cal N}$ is diagonal, however, we note a symmetry (based on a diagonal matrix $\bf Q$)
\begin{eqnarray}
\alpha \rightarrow \alpha {\bf Q} \nonumber \\
\beta \rightarrow {\bf Q}^{-1} \beta \nonumber \\
\alpha \beta - z \beta \alpha \rightarrow \alpha {\bf Q}  {\bf Q}^{-1} \beta - z  {\bf Q}^{-1} \beta \alpha  {\bf Q}  = 1
\end{eqnarray}
such that the modified $\alpha$ and $\beta$ are transposes of each other. The eigenvalues of $\beta \alpha$ are, however, unaffected. The eigenvalues still sit on the same circle on the complex plane. Some further generalizations of the concepts discussed above can yield interesting deviations from a circle \cite{SatishUnpub}.

Though the $\alpha$ and $\beta=\alpha^T$ are $\lambda_0$-dependent, the original $\theta$-commutator (Equation (1)) is independent of $\lambda_0$, i.e., $\alpha \beta - z \beta \alpha=1$. However, the usual commutator does dependent on $\lambda_0$, through the eigenvalues of $\beta \alpha$. In fact,
\begin{eqnarray}
\alpha \beta - \beta \alpha = \alpha \alpha^T - \alpha^T \alpha = \bigg( 1+(z-1) \beta \alpha \bigg)
\end{eqnarray}
The diagonal elements $q_m$ of the commutator matrix in the usual eigenbasis are $q_m=1+(z-1) s_m$. This can be simply re-written as (for $\: m>0$)
\begin{eqnarray}
q_m = 1+(z-1) s_m = 1+ (z-1) \bigg( 1+z+z^2+...+z^{m-1}+z^m \lambda_0 \bigg)  \nonumber \\
= z^{m-1} \bigg( 1+ (z-1) \lambda_0 \bigg) \nonumber
\end{eqnarray}
Hence, we can write the full matrix form as below
\begin{eqnarray}
					     \alpha \beta - \beta \alpha = \alpha \alpha^T - \alpha^T \alpha = (1+(z-1) \lambda_0) \times \left(\begin{array}{cccccccc} {\bf 1} & \: \: \: 0 &\: \: \:  0 &\: \: \:  0 & \: \: \:  0 & ... & 0\\
					     0 & \: \: \: {\bf z} & \: \: \: 0 & \: \: \:  0 & \: \: \:  0 & ... &  \: \: \:  0 & 0 \\
					     0 & \: \: \: 0 & \: \: \: {\bf z^2} &\: \: \: 0 & \: \: \:  0 & ... &  \: \: \:  0 & 0 \\
					     0 & \: \: \: 0 & \: \: \: 0 & \: \: \: {\bf z^3}& \: \: \:  0 & ... &  \: \: \:  0 & 0\\
					      0 & \: \: \: 0 & \: \: \: 0 &\: \: \: 0 & \: \: \: {\bf z^4}& ... &  \: \: \:  0 & 0\\
					     . \\
					     . \\
					     . \\
					     0 & \: \: \: 0 & \: \: \: 0 &\: \: \:  0 & \: \: \:  0 & ... &  {\bf z^{N-2}} &  \: \: \:  0 \\
					     0 & \: \: \: 0 & \: \: \: 0 &\: \: \:  0 & \: \: \:  0 & ... &  \: \: \:  0 &  {\bf z^{N-1}} \end{array}\right)  \: \: \: \: \: \: \: \: \: \: \: \:  \nonumber \\
					     = (1+(z-1) \lambda_0) \times  diag \left[1,z,z^2,z^3, ... , z^{N-1} \right] = (1+(z-1) \lambda_0) \times \Omega \: \: \: \: \: \: \: \: \: \: \: \: \: \: \: \: \: \: \: \: \: \: \: \: \: \: \: \: \: \: \: \: \: \: \: \: \: \:
\end{eqnarray}
which is a scaled version (with scale $1+(z-1) \lambda_0$) of the usual clock matrix ($\Omega$)\cite{Beau}.

Some additional properties of the $\alpha$ matrices are discussed in Appendix 3. Additionally, we plot the eigenvalue spectrum for the special case $\lambda_0=0$ in Fig. 2.

\begin{figure}[h!]
\caption{Eigenvalue Spectrum}
\centering
\includegraphics[scale=.36]{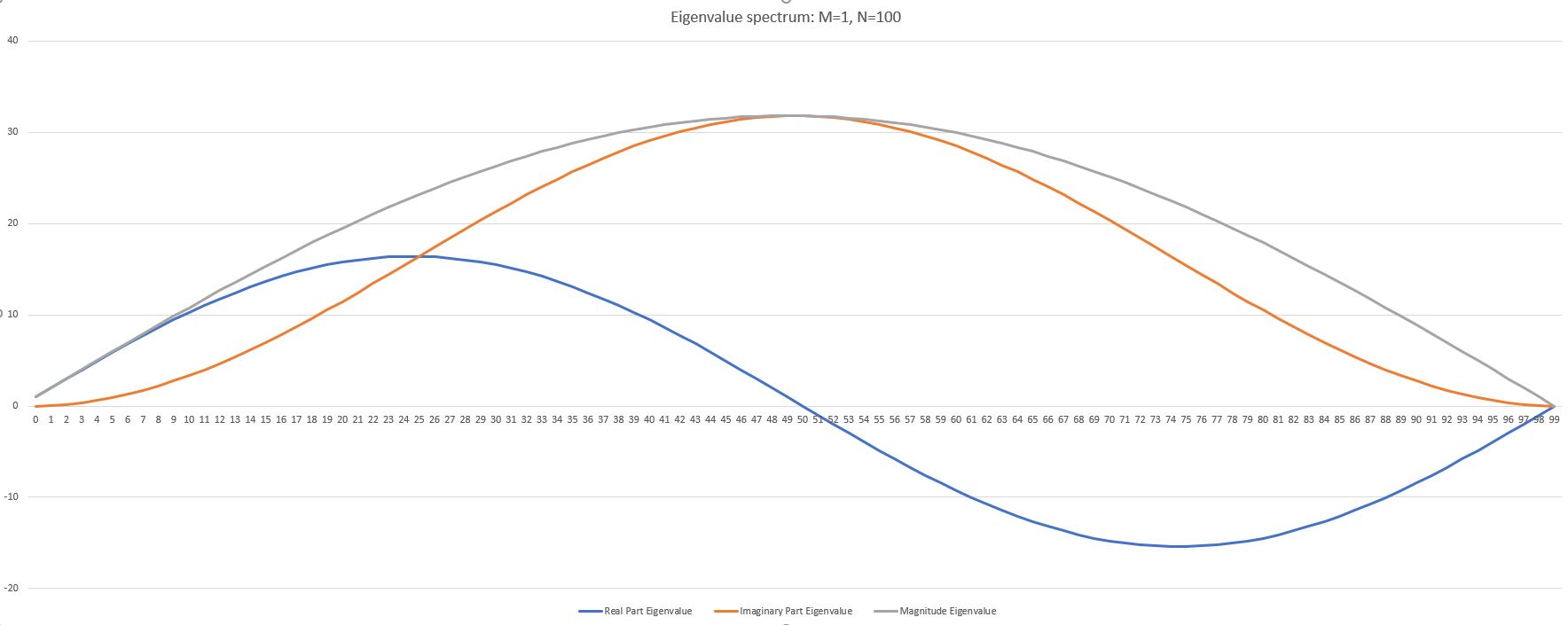}
\includegraphics[scale=0.45]{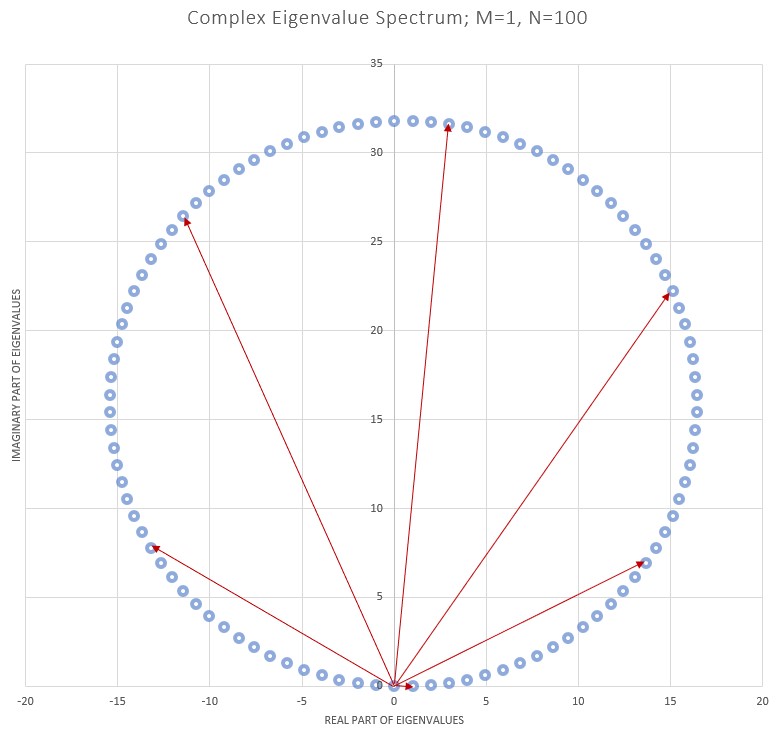}
\end{figure}

\section{Eigenstates of $\alpha$ \: \: :\:  Coherent States}
We can now construct the eigenvalues of $\alpha$. The technique to deduce them is similar to how one deduces eigenvalues for coherent states.

Let's say that $\alpha$ has $N$ eigenstates, denoted by $\ket{\mu}$, with eigenvalue $\mu$. Let's expand $\ket{\mu}$ in the usual eigen-basis, i.e., the eigenstates of the number operator
\begin{eqnarray}
\ket{\mu} = \sum_{i=0}^{N-1} a^{\mu}_i \ket{s_i} \nonumber
\end{eqnarray}

Applying the condition that $\ket{\mu}$ is an eigenstate, we derive,
\begin{eqnarray}
\alpha \ket{\mu} = \mu \ket{\mu} \nonumber \\
\rightarrow \alpha \sum_{i=0}^{N-1} a^{(\mu)}_i \ket{s_i}  = \mu \sum_{i=0}^{N-1} a^{(\mu)}_i \ket{s_i}  \nonumber \\ 
a^{(\mu)}_{i+1}  \sqrt{s_{i+1}} = \mu a_i^{(\mu)} \nonumber \\
\rightarrow a^{(\mu)}_{i+1}   = \frac{\mu a_i^{(\mu)}}{\sqrt{s_{i+1}}}
\end{eqnarray}
The element $a_0$ thus sets the other coefficients. In addition, we also have a normalization condition for the eigenstates. Combining all this, we get
\begin{eqnarray}
a^{(\mu)}_1=\frac{\mu a_0^{(\mu)}}{\sqrt{\lambda_1}} \nonumber \\
a^{(\mu)}_2=\frac{\mu^2 a_0^{(\mu)}}{\sqrt{\lambda_1 \lambda_2}}  \nonumber \\
a^{(\mu)}_3=\frac{\mu^3 a_0^{(\mu)}}{\sqrt{\lambda_1 \lambda_2 \lambda_3}}  \nonumber \\
. \nonumber \\
. \nonumber \\
. \nonumber \\
a^{(\mu)}_{N-1}=\frac{\mu^{N-1} a_0^{(\mu)}}{\sqrt{\lambda_1 \lambda_2...s_{N-}}}  \nonumber \\
\rightarrow  \bigg|a_0^{(\mu)}\bigg|^2 \bigg(1 + \frac{|\mu|^2}{|\lambda_1|}  + \frac{(|\mu|^2)^2 }{|\lambda_1 \lambda_2|} + ...+ \frac{(|\mu|^2)^{N-1} }{|\lambda_1\lambda_2 ... \lambda_{N-1}|} \bigg) = 1
\end{eqnarray}

In addition, we also have, from the circular position of the eigenvalues,
\begin{eqnarray}
a^{(\mu)}_0 \times \sqrt{\lambda_0} = \mu \times a^{\mu}_{N-1} = \mu \times a^{(\mu)}_0 \times \frac{\mu^{N-1}}{\sqrt{\lambda_1 \lambda_2...\lambda_{N-1}}} \nonumber \\
\rightarrow \mu^N = \sqrt{\lambda_0 \lambda_1 ...\lambda_{N-1}}
\end{eqnarray}
In addition, as the eigenvalues $\mu_m$ have the same magnitude $|\mu|= \bigg(|\sqrt{\lambda_0 \lambda_1 ...\lambda_{N-1}}\bigg)^{\frac{1}{N}|}$, $a^{\mu_m}_0$ can be written as $a_0$, independent of the exact value $m$ (this is evident from Equation (13)).

$\mu$ is therefore just the $N^{th}$ root of unity times the $N^{th}$ root of the magnitude of $\sqrt{\lambda_0 \lambda_1 ...\lambda_{N-1}}$. There are $N$ $N^{th}$ roots of unity, all arranged in a circle, hence the eigenvalues of $\alpha$, for $m=0,1,...N-1$, are  $\bigg(\sqrt{\lambda_0 \lambda_1 ...\lambda_{N-1}}\bigg)^{\frac{1}{N}} \times z^m$. We denote these states by $\ket{\mu_m}$ and
\begin{eqnarray}
\ket{\mu_m} = a_0 \: \bigg(  \ket{\lambda_0} + \frac{\mu_m}{\sqrt{\lambda_1}} \ket{\lambda_1} + \frac{\mu_m^2}{\sqrt{\lambda_1 \lambda_2}} \ket{\lambda_2}+ ... + + \frac{\mu_m^{N-1}}{\sqrt{\lambda_1 \lambda_2 ... \lambda_{N-1}}} \ket{\lambda_{N-1}}\bigg) 
\end{eqnarray}

There are, hence, $N$ eigenstates of $\alpha$ in the case $\lambda_0 \ne 0$. These eigenstates are not orthogonal to each other; $\alpha$ is not hermitian (though its determinant is non-zero if $\lambda_0 \ne 0$, so there is no particular requirement that its eigenvectors are orthogonal.

The eigenstates of $\alpha^*$  are the complex conjugates of those of $\alpha$.

If we were to apply this equation at $\lambda_0=0$, we'd discover that all the roots would have to be zero, if they were ordinary complex numbers, since $\mu^N=0$.

When $\lambda_0 \ne 0$, we can also construct eigenstates of $\beta = \alpha^T$. They are
\begin{eqnarray}
\ket{\nu_m} = a_0 \: \bigg(  \ket{\lambda_0} + \frac{\sqrt{\lambda_1} }{\nu_m} \ket{\lambda_1} + \frac{\sqrt{\lambda_1 \lambda_2}}{\nu_m^2 } \ket{\lambda_2}+ ... + + \frac{\sqrt{\lambda_1 \lambda_2 ... \lambda_{N-1}}}{\nu_m^{N-1} } \ket{\lambda_{N-1}}\bigg)  \nonumber \\
\beta \ket{\nu_m} = \nu_m \ket{\nu_m} \: \: \: \: \: \: \: \: \: \: \: \: \: \: \: \: \: \: \: \: \: \: \: \: \: \: \: \: \: \: \: \: \: \: \: \: \:\: \: \: \: \: \: \: \: \: \: \: \: \: \: \: \: \: \: \: \: \: \: \: \: \: \: \: \: \: \: \: \: \: \: \: \: \:
\end{eqnarray}
and, again, $|\nu_m|=\bigg(\sqrt{|\lambda_0 \lambda_1 ...\lambda_{N-1}}\bigg)^{\frac{1}{N}}$. The eigenstates of $\beta$ are a different basis spanning the same space spanned by the eigenvectors of $\alpha$. They are also not orthogonal to each other.

We can also compute the elements of the commutator ${\cal O} = \alpha \beta - \beta \alpha$ in the basis $\ket{\mu_m}$. Note that in the eigen-basis we considered before (of the number operator), this was the well-known clock matrix. To do this, we re-write the operator as ${\cal O} = 1 + (z-1) \beta \alpha = 1+(z-1) {\cal N}$, where ${\cal N} = \beta \alpha$ as defined previously. It helps to write down the scalar product
\begin{eqnarray}
S_{nm}=\bra{\mu_n}\ket{\mu_m} = |a_0|^2 \bigg( 1 + \frac{\mu_n^* \mu_m}{|\lambda_1|}+ ... + \frac{(\mu_n^*)^{N-1} (\mu_m)^{N-1}}{|\lambda_1 \lambda_2 ... \lambda_{N-1}|}\bigg)
\end{eqnarray}
Then,
\begin{eqnarray}
\bra{\mu_n} {{\cal O}} \ket{\mu_m} = \bra{\mu_n} 1+(z-1) {\cal N} \ket{\mu_m} = |a_0^{\mu}|^2 \bigg[ \bigg( 1 + \frac{\mu_n^* \mu_m}{|\lambda_1|}+ ... + \frac{(\mu_n^*)^{N-1} (\mu_m)^{N-1}}{|\lambda_1 \lambda_2 ... \lambda_{N-1}|}\bigg) \nonumber \\
+ (z-1) \bigg( 1 + \frac{\mu_n^* \mu_m \lambda_1}{|\lambda_1|}+ ... + \frac{(\mu_n^*)^{N-1} (\mu_m)^{N-1} \lambda_{N-1}}{|\lambda_1 \lambda_2 ... \lambda_{N-1}|}\bigg) \bigg] \: \: \: \: \: \: \: 
\end{eqnarray}

The diagonal element is (note that $|\mu_m| = \sqrt{|\lambda_0 \lambda_1 ...\lambda_{N-1}|}$ is independent of $m$).
\begin{eqnarray}
\bra{\mu_m} {{\cal O}} \ket{\mu_m} =  |a_0|^2 \bigg[ \bigg( 1 + \frac{|\mu_m|^2}{|\lambda_1|}+ ... + \frac{(|\mu_m|^2)^{N-1}}{|\lambda_1 \lambda_2 ... \lambda_{N-1}|}\bigg) \nonumber \\
+ (z-1) \bigg( 1 + \frac{|\mu_m|^2 \lambda_1}{|\lambda_1|}+ ... + \frac{(|\mu_m|^2)^{N-1} \lambda_{N-1}}{|\lambda_1 \lambda_2 ... \lambda_{N-1}|}\bigg) \bigg] \: \: \: \: \: \: \: \nonumber \\
= 1+ (z-1) \bigg( 1 + \frac{|\mu_m|^2 \lambda_1}{|\lambda_1|}+ ... + \frac{(|\mu_m|^2)^{N-1} \lambda_{N-1}}{|\lambda_1 \lambda_2 ... \lambda_{N-1}|}\bigg) 
\end{eqnarray}

\section{The $N \rightarrow \infty$ (boson) limit}

Let's start with $\lambda_0 \ne 0$ and a finite $N$. The eigenvalues are $..., \frac{\lambda_0-1}{z}, \lambda_0, 1+z \lambda_0, ...$ and there are $N$ of them.

We take $\lambda_0 \rightarrow 0$ first, followed by the limit $N \rightarrow \infty$, which implies $\theta \rightarrow 0, z \rightarrow 1$. The center of the circle would be found at $\frac{1}{2}+i \frac{N}{2 \pi}$ and the radius would tend to $\frac{N}{2 \pi} \sqrt{1 + \frac{4 \pi^2}{N^2} \lambda_0^2} \rightarrow \frac{N}{2 \pi}$. The part of the circle tangent to the real axis would represent the usual bosonic ladder of states. However, there would be negative and positive eigenvalues. We'd have the eigenstates $...-2, -1, 0, 1, 2, ...$, where the states to the left and right of $0$ would be separated - the $\alpha$ and $\beta$ operators would not allow traversal across the $0$ state, since $\alpha \ket{0}=0$ and $\beta \ket{N-1}=0$. The positive eigenvalues correspond to the ones expected for a bosonic oscillator.

On the other hand, if we keep $\lambda_0 \ne 0$ and then take the limit $N\rightarrow \infty$, we'd get a sequence of states $..., \lambda_0-1, \lambda_0, \lambda_0+1, ...$. The center of the circle would be still at $\frac{1}{2}+i \frac{N}{2 \pi}$, which would be infinitely far away along the direction of the imaginary axis. The states would come to lie on a straight, horizontal line at $\lambda_0$ on the complex eigenvalue plane.

\section{A visual model}

A visual model, inspired by the "fuzzy" sphere description of quantized angular momentum states, comes from the realization that for $N=2$, we can establish the following correspondence, i.e.,
\begin{eqnarray}
[ \alpha, \alpha^{\dagger} ] = 2 J_z \: \: \: \: \: \: \: \nonumber \\
J_z = \frac{1}{2} \left(\begin{array}{cc} {\bf 1} & 0\\
0 & -1 \end{array}\right) \: , \: J_x = \frac{\alpha +\alpha^{\dagger}}{2} = \frac{1}{2} \left(\begin{array}{cc} 0 & {\bf 1}\\
{\bf 1} & 0 \end{array}\right) \: , \: J_y = \frac{\alpha -\alpha^{\dagger}}{2i} = \frac{1}{2} \left(\begin{array}{cc} 0 & {\bf -i}\\
{\bf i} & 0 \end{array}\right) \nonumber \\
\left[ J_x, J_y \right]=i J_z \: , \: \left[ J_y, J_z \right ]= i J_x \: , \: \left[ J_z, J_x \right]=i J_y 
\end{eqnarray}

The $\vec J$'s can now be interpreted as coordinates in three dimensions, with $\vec x = \kappa \vec J$, with $x^2+y^2+z^2=R^2$ being the radius of a sphere. In that case, since $\vec J^2=j(j+1)$, this means
\begin{eqnarray}
\kappa^2 = \frac{R^2}{j(j+1)}
\end{eqnarray}
The commutators between the coordinates are
\begin{eqnarray}
\left[ x_i , x_j \right] = i \kappa \: \epsilon_{ijk} x_k \nonumber
\end{eqnarray}
This is a "fuzzy" instead of a full sphere because in this particular situation, $J_z$ is allowed to have only two components $\pm 1$ (since $j=\frac{1}{2}$), which restricts the resolution of the x,y,z-coordinates on the sphere. It is possible to generalize this to the full angular momentum algebra for all $j$. For higher spin $j$, we  obtain, since $\kappa \rightarrow 0$ in that limit, a classical, fully resolved sphere.

Incidentally, the Casimir invariant quantity for the spin-$\frac{1}{2}$ particle is 
\begin{eqnarray}
J_x^2+J_y^2+J_z^2=\frac{3}{4} \nonumber 
\end{eqnarray}
However, as will be explained in a future paper, these operators (for spin-$\frac{1}{2}$ operators, i.e., $N=2$) are also consistent with the equation 
\begin{eqnarray}
\frac{1}{2}(J_x^2+J_y^2)^2+\frac{1}{2} J_z^2 = \frac{1}{4} \nonumber 
\end{eqnarray}

For $N=3$, we obtain, again
\begin{eqnarray}
[ \alpha, \alpha^{\dagger} ] = 2 J_z \: \: \: \: \: \: \: \nonumber \\
J_z = \frac{1}{2} \left(\begin{array}{ccc} {\bf 1} & 0 & 0\\
0 & {\bf |1+z|-1} & 0 \\
0 & 0 & {\bf -|1+z|}  \end{array}\right) \: , \: J_x = \frac{\alpha +\alpha^{\dagger}}{2} = \frac{1}{2} \left(\begin{array}{ccc} 0 & {\bf 1} & 0\\
{\bf 1} & 0 & {\bf \sqrt{1+z}} \\
0 & {\bf \sqrt{1+z^{-1}}}  & 0 \end{array}\right) \: , \nonumber \\
\: J_y = \frac{\alpha -\alpha^{\dagger}}{2i} = \frac{i}{2} \left(\begin{array}{ccc} 0 & {\bf -1} & 0\\
{\bf 1} & 0 & -\bf{\sqrt{1+z}} \\ 
0 & \bf{\sqrt{1+z^{-1}}} &  0 \end{array}\right) \nonumber \\
\left[ J_x, J_y \right]=i J_z \: , \: \left[ J_y, J_z \right ]= i 
\left(1 - \frac{|1+z|}{2} \right) \: J_x \: , \: \left[ J_z, J_x \right]=i \left(1 - \frac{|1+z|}{2} \right) \: J_y \: \: \: 
\end{eqnarray}
To simplify the above, we define the rescaled operators 
\begin{eqnarray}
{\cal A}  = \left( 1 - \frac{|1+z|}{2} \right) \rightarrow \frac{1}{2} \: (\: for \: N=3) \nonumber \\
J_x = \sqrt{\cal A }\hat J_x \: \: \: \: \: \: \: \: \: \: \nonumber \\
J_y = \sqrt{\cal A }\hat J_y  \: \: \: \: \: \: \: \: \: \:  \nonumber \\
J_z = {\cal A} \hat J_z  \: \: \: \: \: \: \: \: \: \: 
\end{eqnarray}
such that the operators $\hat J_{x,y,x}$ satisfy the usual angular momentum algebra. In that case, we can write
\begin{eqnarray}
\frac{J_x^2}{{\cal A}}+\frac{J_y^2}{{\cal A}}+\frac{J_z^2}{{\cal A}^2}=j(j+1)
\end{eqnarray}
which would correspond to a ``fuzzy'' ellipsoid, with $j=1$.

Unfortunately, while this is illustrative, it does not appear to generalize to higher values of $N$ in quite the same way.

We note, in passing that the above equation is consistent with
\begin{eqnarray}
 (J_x^2+J_y^2)^2 \: \sin^2{\frac{\pi}{6}} \:+ J_z^2 \: \cos^2{\frac{\pi}{6}} \: = \frac{1}{4} \nonumber 
\end{eqnarray}
The origin of this will, as mentioned earlier, be discussed in a forthcoming paper.

\section{Conclusions}
We have studied in detail the Hilbert space of an interpolating algebra, that allow us to explore the eigenvalue spectrum at fermion and boson limits as well as between. We have also traced the evolution of the geometrical properties of these eigenstates from each limit to the other. We have then provided an elementary visual mapping, akin to the fuzzy sphere for spin-$\frac{1}{2}$ particles for two particular cases. Generalizations of this will be considered in a future article \cite{Satish2}.

\section{Acknowledgments}

SR acknowledges the hospitality and intellectual stimulation of the Rutgers Department of Physics \& Astronomy and the NHETC at Rutgers. Many parts of this work benefited from very useful advice and suggestions from Professor Scott Thomas. He also acknowledges the collaborative atmosphere provided at the ITP, Santa Barbara.

\section{Appendix 1}

If we start with 
$\alpha \beta - z \beta \alpha = 1$, then the re-definition below
\begin{eqnarray}
\alpha = \frac{i}{\sqrt{z}} \hat \alpha \: \: ; \: \: 
\beta = \frac{i}{\sqrt{z}} \hat \beta \nonumber
\end{eqnarray}
implies
\begin{eqnarray}
\rightarrow \: \hat \beta \hat \alpha - \frac{1}{z} \hat \alpha \hat \beta = 1
\end{eqnarray}
For bosons, the number operator $\cal N= \beta \alpha$ switches sign when we make the transformation, while for fermions, the number operator does not suffer a change of sign. 

In fact the mapping
\begin{eqnarray}
\alpha \rightarrow \frac{i}{\sqrt{z}} \beta \: \: ; \: \: 
\beta \rightarrow \frac{i}{\sqrt{z}} \alpha \nonumber \\
\theta \rightarrow -\theta
\end{eqnarray}
is a symmetry of the algebra - it represents a background automorphism of the algebra.

\section{Appendix 2}
Consider the distance of the $n^{th}$ eigenvalue from the center at $\frac{1}{1-z}$. This is
\begin{eqnarray}
{\cal D}_n=|1+z+z^2+z^3+...+z^{n-1}+z^n \lambda_0 - \frac{1}{1-z}| \nonumber \\
=\frac{| (1+z+z^2+z^3+...+z^{n-1}+z^n \lambda_0) - (z+z^2+z^3+...+z^{n-1}+z^n \lambda_0)  -1|}{|1-z|} \nonumber \\
=| \frac{1+(z-1) \lambda_0}{1-z} |  \: \: \: \: \: \: \: \: \: \: \: \: \: \: \: \: \: \:  \: \: \: \: \: \: \: \: \:  \: \: \: \: \: \: \: \: \:  \: \: \: \: \: \: \: \: \: 
\end{eqnarray}
as in the text.

A simple way to see that the points lie on a circle is to write the $n^{th}$ eigenvalue as
\begin{eqnarray}
\lambda_n = 1+z+z^2+z^3+...+z^{n-1}+z^n \lambda_0 = \frac{1-z^n}{1-z}+z^n \lambda_0  \nonumber \\
= \frac{1}{1-z} + z^n \frac{1+(z-1) \lambda_0}{z-1}
\end{eqnarray}
which writes the complex eigenvalue $\lambda_n$ as a complex coordinate for the center, i.e., $\frac{1}{1-z}$ plus a progressively rotated (by $z^n$) radius complex vector $\frac{1+(z-1) \lambda_0}{z-1}$.

\section{Appendix 3}

Passing $\alpha$ and $\alpha^{\dagger}$ through diagonal matrices, we get
\begin{eqnarray}
\alpha \: . \:  \left(\begin{array}{cccccccc} {\bf A_0} & \: \: \: 0 &\: \: \:  0 &\: \: \:  0 &\: \: \:  0 & ... & \: \: \:  0 &0\\
					     0 & \: \: \: {\bf A_1} & \: \: \: 0 & \: \: \:  0 & \: \: \:  0 & ... & \: \: \:  0 &0 \\
					     0 & \: \: \: 0 & \: \: \: {\bf A_2} &\: \: \: 0 & \: \: \:  0 &... & \: \: \:  0 &0 \\
					     0 & \: \: \: 0 & \: \: \: 0 &\: \: \: {\bf A_3} & \: \: \: 0 &... & \: \: \:  0 &0 \\
					     . \\
					     0 & \: \: \: 0 & \: \: \: 0 &\: \: \:  0 & \: \: \:  0 & ... &\: \: \:  {\bf A_{N-2}}  &0  \\
					    0 & \: \: \: 0 & \: \: \: 0 &\: \: \:  0 & \: \: \:  0 & ... &\: \: \:  0 & {\bf A_{N-1}} \end{array}\right)  \: \: \: \: \: \: \: \: \: \: \: \: \: \: \: \: \: \:  \: \: \: \: \: \: \: \: \:  \: \: \: \: \: \: \: \: \:  \: \: \: \: \: \: \: \: \: \nonumber \\
					  \: \: \: \: \: \: \: \: \:   = 
					    \left(\begin{array}{cccccccc} {\bf A_1} & \: \: \: 0 &\: \: \:  0 &\: \: \:  0 &\: \: \:  0 & ... & \: \: \:  0 &0\\
					     0 & \: \: \: {\bf A_2} & \: \: \: 0 & \: \: \:  0 & \: \: \:  0 & ... & \: \: \:  0 &0 \\
					     0 & \: \: \: 0 & \: \: \: {\bf A_3} &\: \: \: 0 & \: \: \:  0 &... & \: \: \:  0 &0 \\
					     0 & \: \: \: 0 & \: \: \: 0 &\: \: \: {\bf A_4} & \: \: \: 0 &... & \: \: \:  0 &0 \\
					     . \\
					     0 & \: \: \: 0 & \: \: \: 0 &\: \: \:  0 & \: \: \:  0 & ... &\: \: \:  {\bf A_{N-1}}  &0  \\
					    0 & \: \: \: 0 & \: \: \: 0 &\: \: \:  0 & \: \: \:  0 & ... &\: \: \:  0 & {\bf A_0} \end{array}\right) \: . \: \alpha 
\end{eqnarray}
and
\begin{eqnarray}
\alpha^{\dagger} \: . \:  \left(\begin{array}{cccccccc} {\bf A_0} & \: \: \: 0 &\: \: \:  0 &\: \: \:  0 &\: \: \:  0 & ... & \: \: \:  0 &0\\
					     0 & \: \: \: {\bf A_1} & \: \: \: 0 & \: \: \:  0 & \: \: \:  0 & ... & \: \: \:  0 &0 \\
					     0 & \: \: \: 0 & \: \: \: {\bf A_2} &\: \: \: 0 & \: \: \:  0 &... & \: \: \:  0 &0 \\
					     0 & \: \: \: 0 & \: \: \: 0 &\: \: \: {\bf A_3} & \: \: \: 0 &... & \: \: \:  0 &0 \\
					     . \\
					     0 & \: \: \: 0 & \: \: \: 0 &\: \: \:  0 & \: \: \:  0 & ... &\: \: \:  {\bf A_{N-2}}  &0  \\
					    0 & \: \: \: 0 & \: \: \: 0 &\: \: \:  0 & \: \: \:  0 & ... &\: \: \:  0 & {\bf A_{N-1}} \end{array}\right)  \: \: \: \: \: \: \: \: \: \: \: \: \: \: \: \: \: \:  \: \: \: \: \: \: \: \: \:  \: \: \: \: \: \: \: \: \:  \: \: \: \: \: \: \: \: \: \nonumber \\
					  \: \: \: \: \: \: \: \: \:   = 
					    \left(\begin{array}{cccccccc} {\bf A_{N-1}} & \: \: \: 0 &\: \: \:  0 &\: \: \:  0 &\: \: \:  0 & ... & \: \: \:  0 &0\\
					     0 & \: \: \: {\bf A_0} & \: \: \: 0 & \: \: \:  0 & \: \: \:  0 & ... & \: \: \:  0 &0 \\
					     0 & \: \: \: 0 & \: \: \: {\bf A_1} &\: \: \: 0 & \: \: \:  0 &... & \: \: \:  0 &0 \\
					     0 & \: \: \: 0 & \: \: \: 0 &\: \: \: {\bf A_2} & \: \: \: 0 &... & \: \: \:  0 &0 \\
					     . \\
					     0 & \: \: \: 0 & \: \: \: 0 &\: \: \:  0 & \: \: \:  0 & ... &\: \: \:  {\bf A_{N-3}}  &0  \\
					    0 & \: \: \: 0 & \: \: \: 0 &\: \: \:  0 & \: \: \:  0 & ... &\: \: \:  0 & {\bf A_{N-2}} \end{array}\right) \: . \: \alpha^{\dagger} 
\end{eqnarray}
Hnce, $\alpha$ or $\alpha^{\dagger}$ are similiar in structure to the Shift matrices of the Clock/Shift algebra. $\alpha^{\dagger}$ produces a diagonal shift-down, while $\alpha$ produces a diagonal shift-up in a diagonal matrix.

\section{Appendix 4}

What's the most general set of operators that commute with ${\cal N}$? Are there any others than $\cal N$ itself and functions of $\cal N$?

We start with the following four relations between $\alpha$, $\beta$ and $\cal N$.
\begin{eqnarray}
\alpha {\cal N} = (1 + z {\cal N}) \: \alpha \nonumber \\
{\cal N} \alpha= \alpha \frac{\: {\cal N}-1}{z}  \nonumber \\
 {\cal N} \beta = \beta \: (1 + z {\cal N}) \nonumber \\
\beta {\cal N} =  \frac{\: {\cal N}-1}{z} \: \beta
\end{eqnarray}

Some powers can also be computed, i.e.,
\begin{eqnarray}
\alpha^m {\cal N} = (1 + z+z^2 + ...+ z^m {\cal N}) \: \alpha^m \nonumber \\
{\cal N} \beta^n= \beta^n (1 + z+z^2 + ...+ z^n {\cal N}) \:  \nonumber \\
\beta^n {\cal N} = \frac{{\cal N} - (1+z+z^2+...+z^{n-1})}{z^n} \beta^n
\end{eqnarray}

The most general operator made up of $\alpha$ and $\beta$ can be written as ${\cal X} = C_{nm} \beta^n \alpha^m$. Any different ordering of the $\alpha$'s and $\beta$'s can be brought to this form by using the basic commutator. We require that it commute with $\cal N$, i.e.,
\begin{eqnarray}
\bigg[\beta^n \alpha^m, {\cal N} \bigg] \nonumber \\
 = \bigg( (1+z+...+z^{m-1}) - z^{m-n} (1+z+...+z^{n-1}) - {\cal N} (1 - z^{m-n}) \bigg) \beta^n \alpha^m 
\end{eqnarray}
For the commutator with ${\cal X}$ to give $0$, we must have (by inspection of the above), $m=n$. Hence the only possible operator that commutes with $\cal N$ is ${\cal X} = C_n \beta^n \alpha^n$. These polynomials can {\underline all} be written in terms of $\cal N$, as is evident from
\begin{eqnarray}
\beta^0 \alpha^0 = {\cal I} \nonumber \\
\beta^1 \alpha^1 = {\cal N} \nonumber \\
\beta^2 \alpha^2 = \frac{{\cal N}-1}{z} \: {\cal N} \nonumber \\
\beta^3 \alpha^3 = \frac{{\cal N}-(1+z)}{z} \:  \frac{{\cal N}-1}{z} \: {\cal N} \nonumber \\
... \nonumber \\
\beta^n \alpha^n = \frac{{\cal N}-(z^0+z^1+..+z^{n-2})}{z} \:  ...\frac{{\cal N}-(1+z)}{z} \:  \frac{{\cal N}-1}{z} \: {\cal N}
\end{eqnarray}

If one considers all combinations of polynomials of the type $\beta^n \alpha^m$, there are $N^2$ of them, however, only $N$ commute with $\cal N$. The rest $N^2-N$ operators do not commute with $\cal N$ and are not diagonal in the canonical basis.

\end{document}